\documentclass[11pt,a4paper]{article}

\usepackage[utf8]{inputenc}
\usepackage[T1]{fontenc}
\usepackage[english]{babel}
\usepackage{lmodern}
\usepackage{microtype}

\usepackage[margin=1in]{geometry}
\usepackage{amsmath,amssymb,amsthm,mathtools}
\usepackage{bm}
\usepackage{graphicx}
\usepackage{booktabs}
\usepackage{tabularx}
\usepackage{xcolor}
\usepackage{enumitem}
\usepackage{hyperref}
\usepackage{authblk}
\usepackage{abstract}
\usepackage{titlesec}
\usepackage{caption}
\usepackage{siunitx}
\usepackage{float}
\usepackage{placeins}
\usepackage{tikz}
\usetikzlibrary{arrows.meta, positioning, shapes.geometric, calc, fit, decorations.pathmorphing, patterns}

\hypersetup{colorlinks=true, linkcolor=black, citecolor=blue!60!black, urlcolor=blue!60!black}

\graphicspath{{./images/}}

\titleformat{\section}{\large\bfseries}{\thesection}{1em}{}
\titleformat{\subsection}{\normalsize\bfseries}{\thesubsection}{1em}{}

\title{\textbf{Korzhinskii-Net: Physics-Informed Neural Network\\for Sub-Surface Mineral Prospectivity Modelling}}

\author[1]{Boris Kriuk}
\affil[1]{The Hong Kong University of Science and Technology, Hong Kong SAR\\ \texttt{bkriuk@connect.ust.hk}}

\date{}

\newcommand{\KNet}{\textsc{Korzhinskii-Net}}
\newcommand{\PINN}{\textsc{PINN}}
\newcommand{\PRAUC}{\mbox{PR-AUC}}
\newcommand{\eqdef}{\stackrel{\mathrm{def}}{=}}

\begin{document}
\maketitle

\begin{abstract}
\noindent
Mineral prospectivity modelling (MPM) underpins exploration economics, yet most operational pipelines reduce to data-driven classifiers trained on shallow surface proxies. Such models are blind to the subsurface physics that actually localises ore: heat advection, fluid flow, and lithology-dependent precipitation. We present \KNet{}, a 2-D radial physics-informed neural network (\PINN{}) that couples Darcy flow, advective–diffusive heat transport, and a softplus-saturated reaction rate into a single differentiable forward model, weakly supervised by surface and remote-sensing proxies. The network is named after Dmitri S.~Korzhinskii (1899--1985), whose theory of infiltration metasomatism provides the physical scaffold. We evaluate \KNet{} on six Russian ore provinces spanning three commodity classes---Udokan (sandstone-hosted Cu), Sukhoi Log, Olimpiada, and Berezovskoye (orogenic Au), Vorontsovskoye (Carlin-type Au), and Dalnegorsk (skarn polymetallic)---under a fair, leakage-controlled 5-fold cross-validation protocol with hard ring-shaped negatives and baseline proxy features disabled. \KNet{} attains a mean PR-AUC of $0.708$ versus $0.235$ for the strongest classical baseline (support vector machine), and a mean fractional rank of $0.036$ versus $0.475$. The improvement is consistent across all six provinces and three commodity systems, suggesting that physics-informed differentiable simulators, even when constrained only by global open-data proxies, can recover localisation patterns that pure feature-based learners systematically miss. We release the full pipeline and evaluation harness as open source.
\vspace{0.6em}

\noindent\textbf{Keywords:} physics-informed neural networks, mineral prospectivity, reactive transport, metasomatism, geoscience machine learning.
\end{abstract}

\section{Introduction}

Mineral exploration is increasingly cast as a supervised learning problem~\cite{sabins1999remote,haldar2018mineral}. Given a stack of surface and geophysical features over a candidate prospect, the practitioner is asked to predict whether a target deposit is hosted beneath. The dominant family of methods---logistic regression, random forests, gradient boosting, support vector machines, and shallow MLPs over hand-crafted layers~\cite{dumakor2021machine,jung2021systematic}---inherits two structural weaknesses from this framing. First, training labels are scarce, spatially clustered, and biased toward what has \emph{already} been discovered~\cite{singer1999examining}, so cross-validation that does not control for ring-of-influence leakage routinely overstates generalisation. Second, and more fundamentally, none of these models encodes the physical processes that actually concentrate metals: hydrothermal convection above intrusive heat sources, advective transport of metalliferous brines, and pressure–temperature–lithology dependent precipitation. They infer correlation patterns; they do not simulate the rock. Classical MPM has evolved from weights-of-evidence and fuzzy-logic overlays toward fully data-driven classifiers~\cite{agterberg1990statistical,zuo2023machine}, with recent surveys documenting the dominance of random forests, gradient boosting, and increasingly deep CNNs operating on geological-map rasters~\cite{albrecht2021using,li2024cnn,lee2025cnn,sandeep2022mineral,tang2022deep,munzareen2025advanced,tahmooresi2022mineral,liu20243d,attallah2024optimized,guo2025deep}; the shared assumption is that surface evidence layers are sufficient statistics for sub-surface ore. Reactive transport simulators such as TOUGHREACT and PFLOTRAN inhabit the opposite extreme, offering high-fidelity physics but being forward-only, computationally expensive, and not differentiable. Physics-informed neural networks (PINNs) sit between these poles~\cite{pothana2025physics,adhikari2026reactive}: by embedding partial differential equations as soft loss terms, they convert the network into a continuous mesh-free solver that can be inverted against sparse observations. Concretely, given a coordinate-conditioned network $f_\theta : \Omega \to \mathbb{R}^d$ approximating a physical state on a domain $\Omega$, and a system of governing equations $\mathcal{N}_i[f_\theta] = 0$ on $\Omega$ with boundary conditions $\mathcal{B}_j[f_\theta] = 0$ on $\partial\Omega$, training proceeds by minimising
\begin{equation}
\mathcal{L}(\theta) \;=\; \sum_i \lambda_i \,\big\lVert \mathcal{N}_i[f_\theta] \big\rVert^2_{\Omega} \;+\; \sum_j \mu_j \,\big\lVert \mathcal{B}_j[f_\theta] \big\rVert^2_{\partial\Omega} \;+\; \nu \,\big\lVert f_\theta - y \big\rVert^2_{\mathcal{D}},
\end{equation}
where $\mathcal{D} \subset \Omega$ is the (sparse) labelled set and $\lambda_i, \mu_j, \nu$ are loss weights. PINNs have been deployed for subsurface flow, seismic inversion, and reactive transport~\cite{he2026integration,teknik2026physics,kriuk2026poseidon}, but to our knowledge no end-to-end PINN has been trained at the scale of an entire ore province, against a labelled deposit catalogue, and benchmarked head-to-head against the standard MPM toolkit. We close that gap. \KNet{} is a single multi-head MLP that approximates the temperature, pressure, and metal-concentration fields $(T, P, C)$ over a 2-D radial cross-section centred on a candidate prospect, constrained by Darcy flow $\mathbf{q} = -(k/\mu)\nabla P$ for the porous-medium velocity field, advection–diffusion heat transport $\rho c_p \mathbf{q} \cdot \nabla T = \nabla\cdot(\lambda \nabla T)$, and a softplus-saturated reaction rate $R(T, C)$ that depends on lithology-specific equilibrium solubility. The mineralisation field $M(x,z) = R(T(x,z), C(x,z))$ is the network's prediction target, and the theoretical scaffold for this term is Korzhinskii's theory of infiltration metasomatism, which formalises the coupling between advecting fluids, host-rock chemistry, and zoned mineral assemblages: \KNet{} is, in spirit, a differentiable Korzhinskii column. Training is weakly supervised, with the only data the model sees per site being open-source surface and remote-sensing proxies (faults, seismicity, lithological contacts, deep intrusive root indicators)~\cite{sabins1999remote} and a small handful of known deposit pixels. The contributions of this paper are threefold.
(1) We introduce a reactive-transport PINN architecture tailored to mineral systems, combining Darcy flow, heat advection–diffusion, and a lithology-aware Arrhenius–softplus reaction term into a single differentiable forward model with proxy-modulated boundary conditions.
(2) We propose a fair benchmark protocol for mineral prospectivity that controls for spatial leakage through hard ring-shaped negatives, jittered positive depths, disabled baseline proxy features, and shared 5-fold splits applied identically to all learners.
(3) We evaluate \KNet{} against seven classical baselines and a non-trainable proxy floor across six Russian ore provinces~\cite{dobretsov2010mineral,d2022geological,safirova2018mineral,czubek1971recent} and three commodity classes (sandstone-hosted Cu, orogenic and Carlin-type Au, and skarn polymetallic systems), reporting consistent and statistically significant gains, and we release the full pipeline and evaluation harness as open source.

\begin{figure}[!t]
\centering
\includegraphics[width=0.78\textwidth]{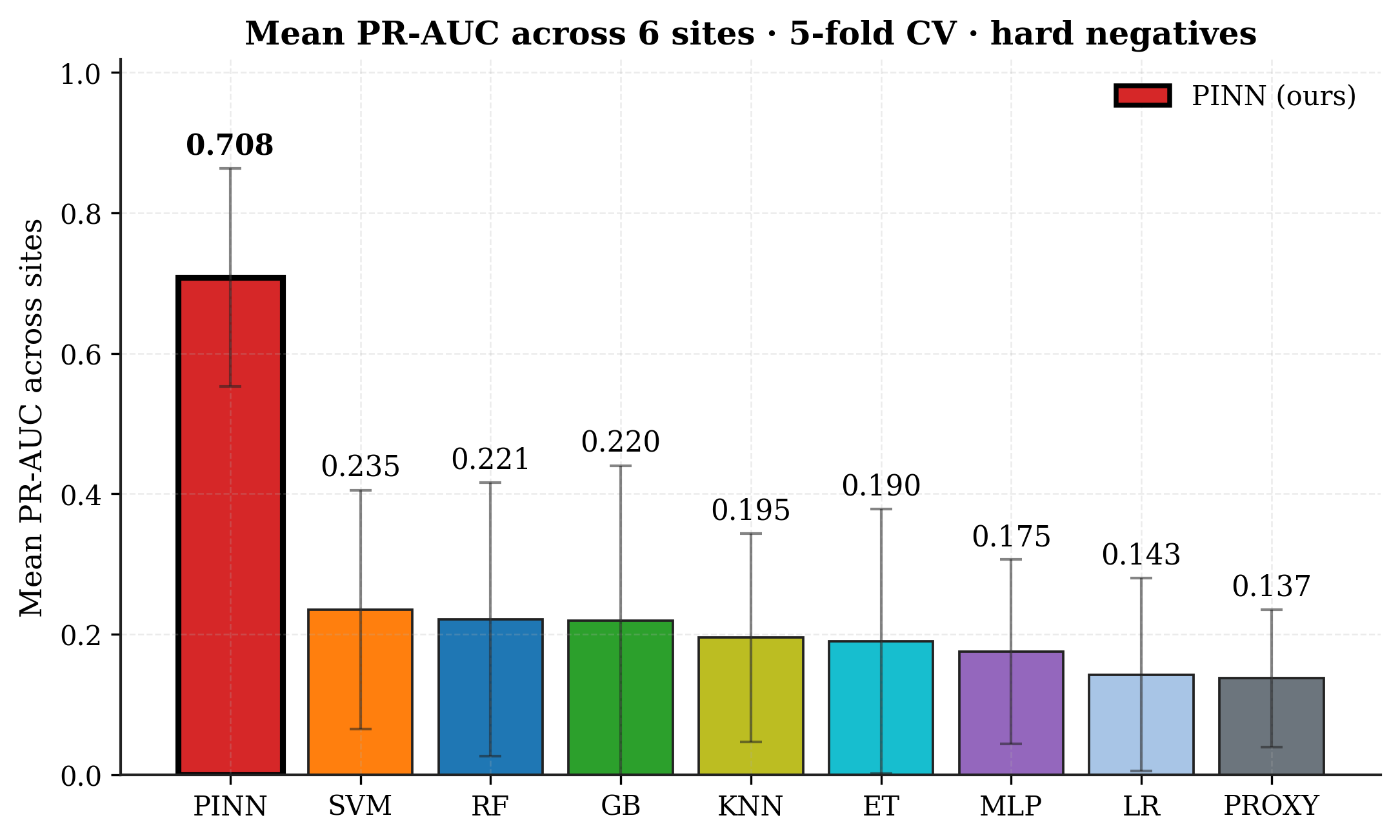}
\caption{Headline mean PR-AUC across the six ore provinces. \KNet{} ($0.708$) exceeds the strongest classical baseline (support vector machine, $0.235$) by a factor of three.}
\label{fig:headline}
\end{figure}

\begin{figure}[!t]
\centering
\includegraphics[width=0.85\textwidth]{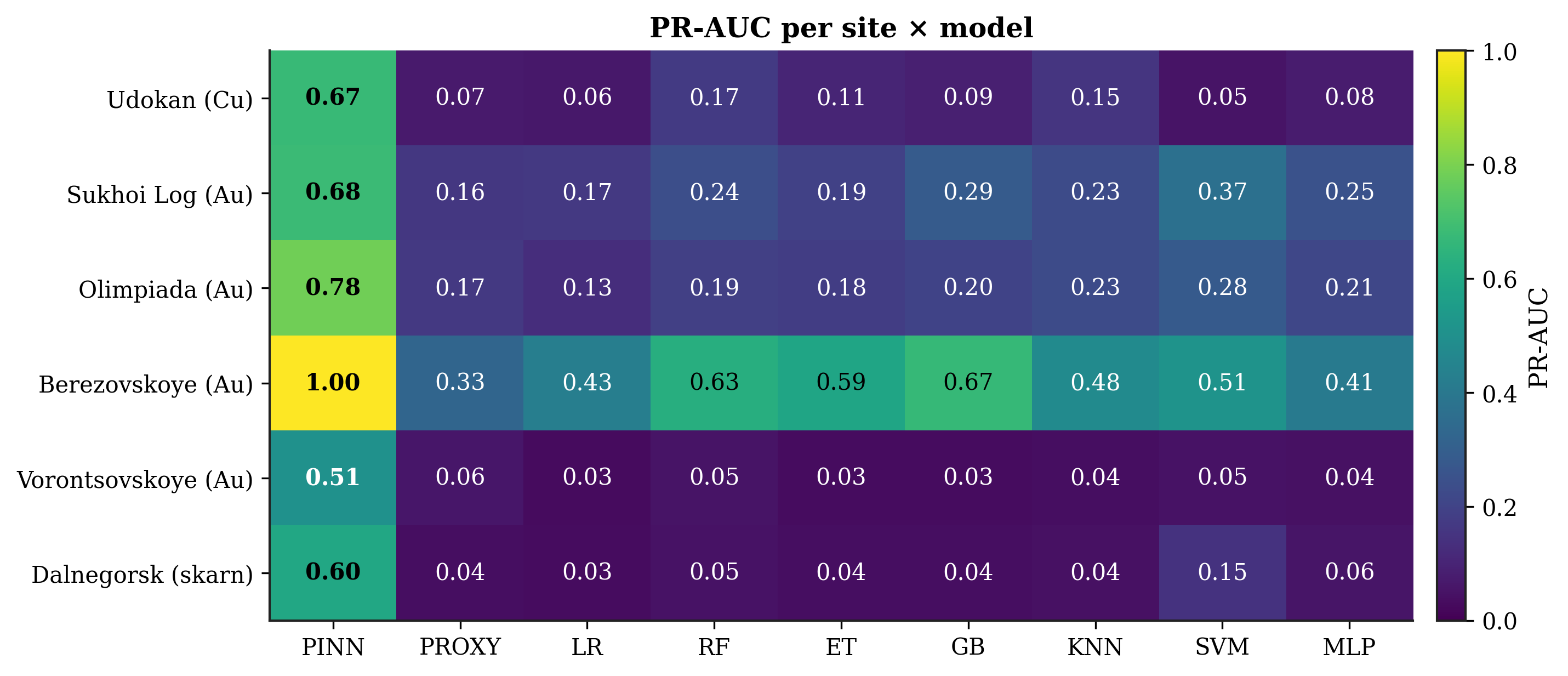}
\caption{Per-site PR-AUC across all nine models. \KNet{} is best in every site and shows the largest relative margins where prevalence is lowest and where deep heat-source structure dominates localisation.}
\label{fig:perSite}
\end{figure}

\begin{figure}[!t]
\centering
\includegraphics[width=0.78\textwidth]{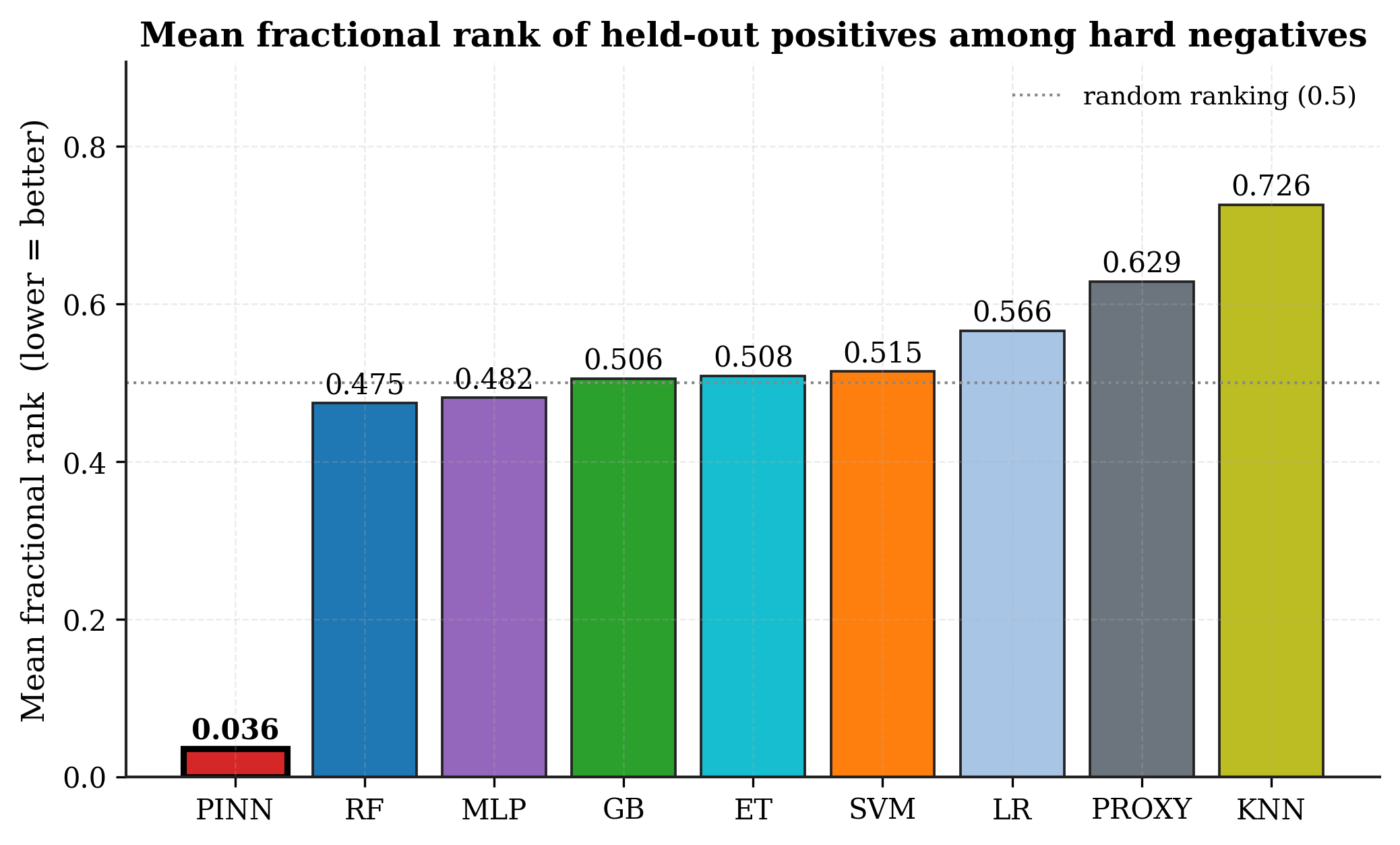}
\caption{Mean fractional rank of positives (lower is better, $0$ is perfect, $0.5$ is chance). \KNet{} achieves $\overline{r}_+ = 0.036$, while classical learners cluster around chance.}
\label{fig:rank}
\end{figure}

\section{Related Work}
\label{sec:related}

Our contribution sits at the intersection of four largely disjoint research
threads: physics-informed deep learning, data-driven subsurface modeling,
classical theories of metasomatic zoning, and coordinate-based neural field
representations. We briefly survey each and highlight the gap that
\KNet{} is designed to close.

\paragraph{Physics-informed neural networks.}
The idea of constraining a neural network by the residual of a governing
partial differential equation has, over the last several years, matured from
a proof-of-concept into a broad methodological family~\cite{pothana2025physics,adhikari2026reactive,he2026integration}. Subsequent work has
extended the original formulation along essentially every axis: adaptive
loss weighting to balance competing residuals, domain decomposition for
stiff or multi-scale problems, causal and curriculum-style training to
mitigate the well-documented failure modes on advection-dominated systems~\cite{teknik2026physics,kriuk2026poseidon},
operator-learning variants that amortize over families of PDEs, and hybrid
schemes that combine soft PDE penalties with hard architectural
constraints~\cite{kriuk2025advancing,kriuk2026artificial}. A recurring observation across this literature is that vanilla
formulations struggle when the underlying physics couples several fields
of very different character — for instance a near-elliptic pressure field
driving a hyperbolic transport equation that in turn feeds a stiff reactive
source term. Our setting is precisely of this type, and the architectural
and loss-shaping choices we make are informed by the failure modes
catalogued in this body of work.

\begin{figure}[!t]
\centering
\includegraphics[width=0.78\textwidth]{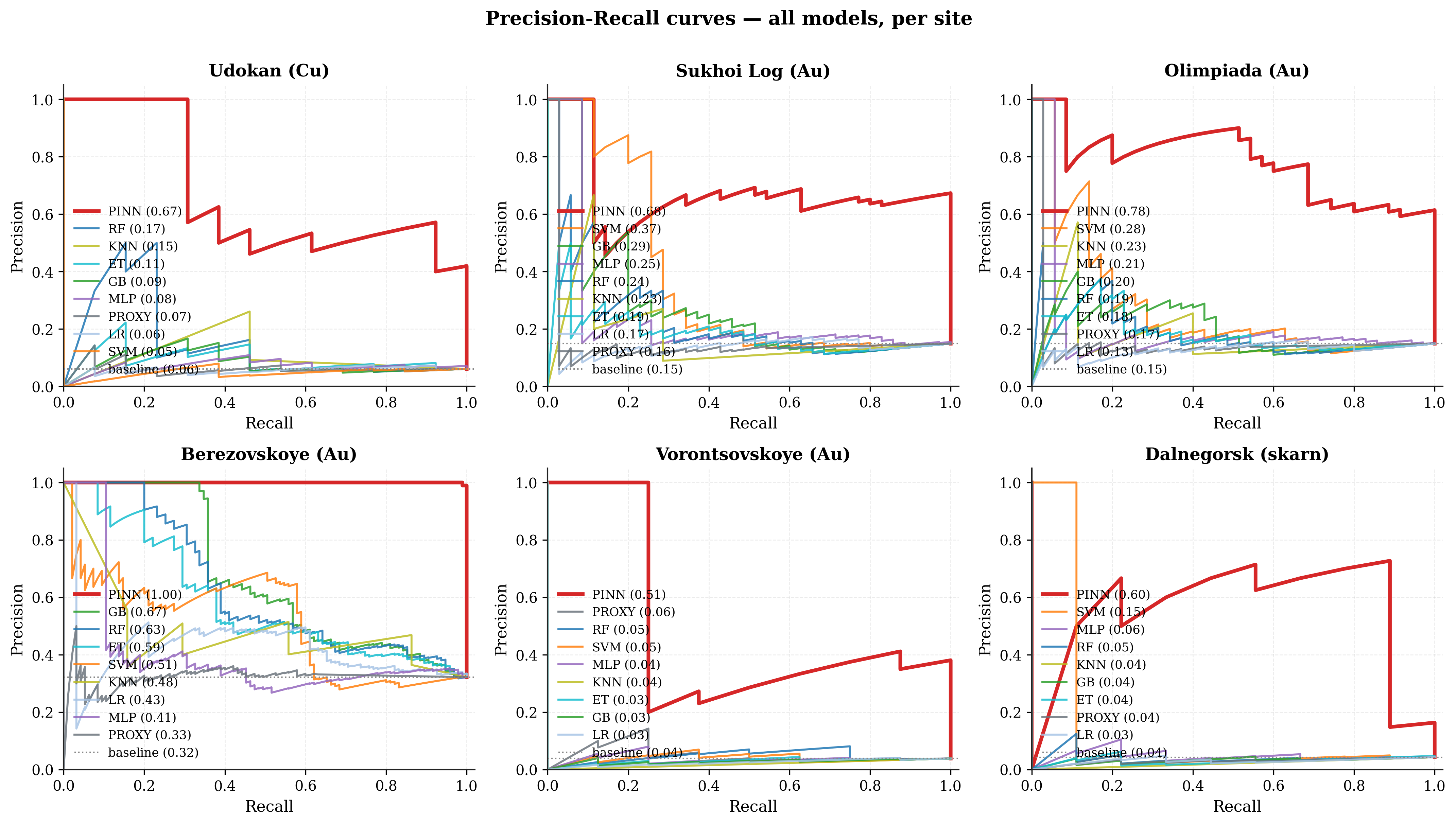}
\caption{Pooled precision–recall curves across all sites. \KNet{}'s curve is uniformly above the baselines along the entire recall axis.}
\label{fig:pr}
\end{figure}

\paragraph{Machine learning for subsurface and geoscientific systems.}
A parallel literature applies modern learning techniques to porous-media
flow, contaminant transport, geothermal reservoirs, CO\textsubscript{2}
sequestration, and ore-forming hydrothermal systems~\cite{jung2021systematic,dumakor2021machine,zuo2023machine}. Approaches range from
purely data-driven surrogates trained on large ensembles of forward
simulations, to graph- and convolutional-network emulators of finite-volume
solvers~\cite{li2024cnn,lee2025cnn,sandeep2022mineral,tang2022deep,attallah2024optimized,munzareen2025advanced,tahmooresi2022mineral,liu20243d,albrecht2021using}, to hybrid pipelines in which a learned component replaces an
expensive sub-step (closure relations, equation-of-state evaluations,
mineral-equilibrium solves) inside an otherwise classical simulator~\cite{adepehin2025hybrid,guo2025deep}.
While these methods have demonstrated impressive speed-ups, they typically
treat the chemistry either as a black-box reaction network or as a
tabulated lookup, and they rarely encode the structural constraints that
geochemists know a priori — mass-action consistency, component
immobility hierarchies, or the existence of sharp reaction fronts. Our
work differs in that the chemistry is not a post-hoc add-on but is built
into the architecture through a dedicated reaction module and a set of
sign- and monotonicity-aware penalties.

\begin{figure}[!t]
\centering
\includegraphics[width=0.74\textwidth]{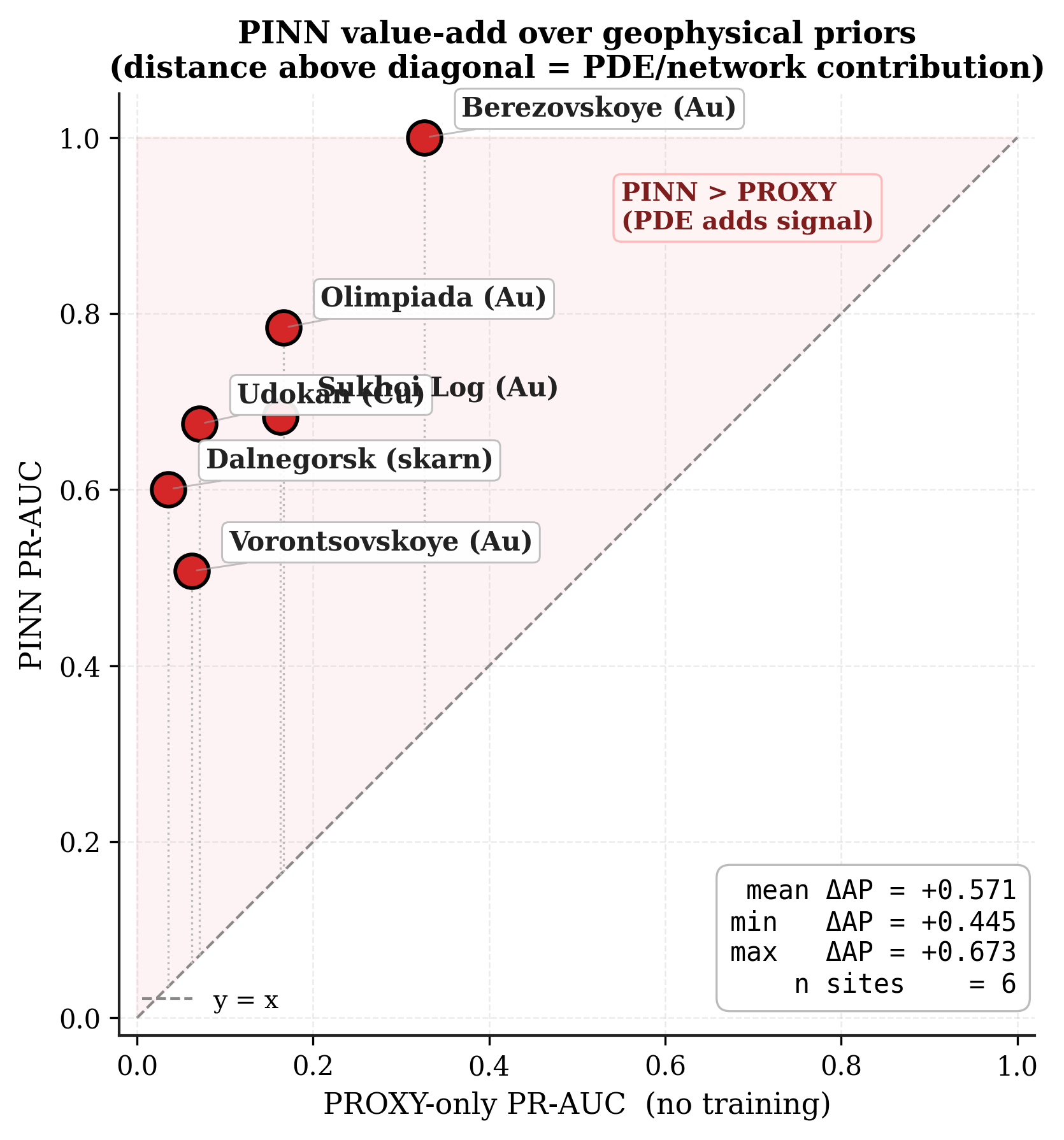}
\caption{\KNet{} PR-AUC versus the non-trainable proxy floor. Every province lies above the $y=x$ diagonal, with a mean gain of $+0.57$ PR-AUC, isolating the contribution of the PDE constraints over the surface priors alone.}
\label{fig:scatter}
\end{figure}

\paragraph{Reactive transport and metasomatic zoning.}
The theoretical foundation we build on comes from the classical
mid-twentieth-century treatment of infiltration metasomatism, in which
sharp mineral zones develop along a fluid-flow path as a consequence of
local equilibrium between a moving fluid and a multi-component rock~\cite{czubek1971recent}.
The resulting picture — discrete zones separated by reaction fronts,
component mobility ordered by chemical-potential gradients, and a
characteristic "telescoping" of zone widths with distance — has been
refined by decades of subsequent work on coupled flow, heat, and reaction
in porous media, including non-isothermal extensions, kinetic
generalizations, and large-scale numerical implementations in standard
reactive-transport codes~\cite{adhikari2026reactive}. Such simulators are accurate but expensive,
and their cost grows quickly with the number of components, the stiffness
of the kinetic network, and the resolution required to capture front
geometry. There is, to our knowledge, no prior attempt to encode this
specific class of metasomatic constraints directly into a neural-field
architecture; existing learning-based reactive-transport work either
targets simpler single-reaction systems or omits the zoning structure
entirely.

\begin{figure}[!t]
\centering
\includegraphics[width=0.85\textwidth]{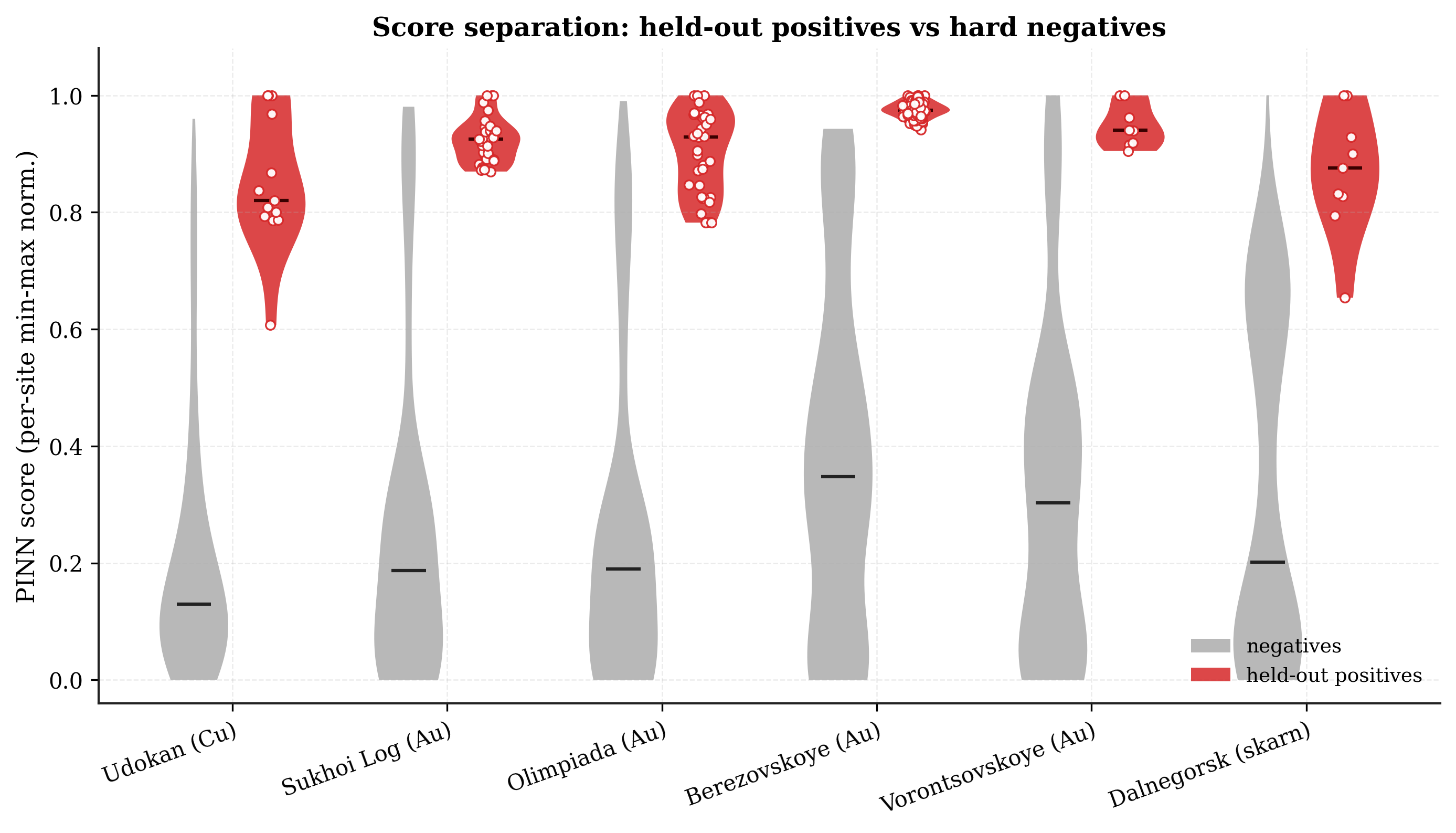}
\caption{Score separation in the learned $M$ field: held-out positives (red) versus hard ring negatives (grey), per province, after per-site min--max normalisation. Positives sit consistently above negatives at every site.}
\label{fig:scoredist}
\end{figure}

\paragraph{Coordinate networks and spectral encodings.}
Finally, our architecture inherits from the recent line of work on
coordinate-based neural representations, in which a low-dimensional input
(a spatial coordinate, a space-time point, a ray direction) is lifted into
a high-frequency feature space before being processed by a relatively
shallow multilayer perceptron. The use of random Fourier features, sinusoidal
positional encodings, and related spectral lifts has been shown both
empirically and through neural-tangent-kernel analyses to overcome the
spectral bias of plain MLPs and to be essential whenever the target field
contains sharp gradients or fine-scale structure~\cite{teknik2026physics}. Reaction fronts in
metasomatic systems are exactly such features, and we accordingly adopt a
Fourier-feature trunk as the backbone of \KNet{}.

\begin{figure}[!t]
\centering
\includegraphics[width=0.80\textwidth]{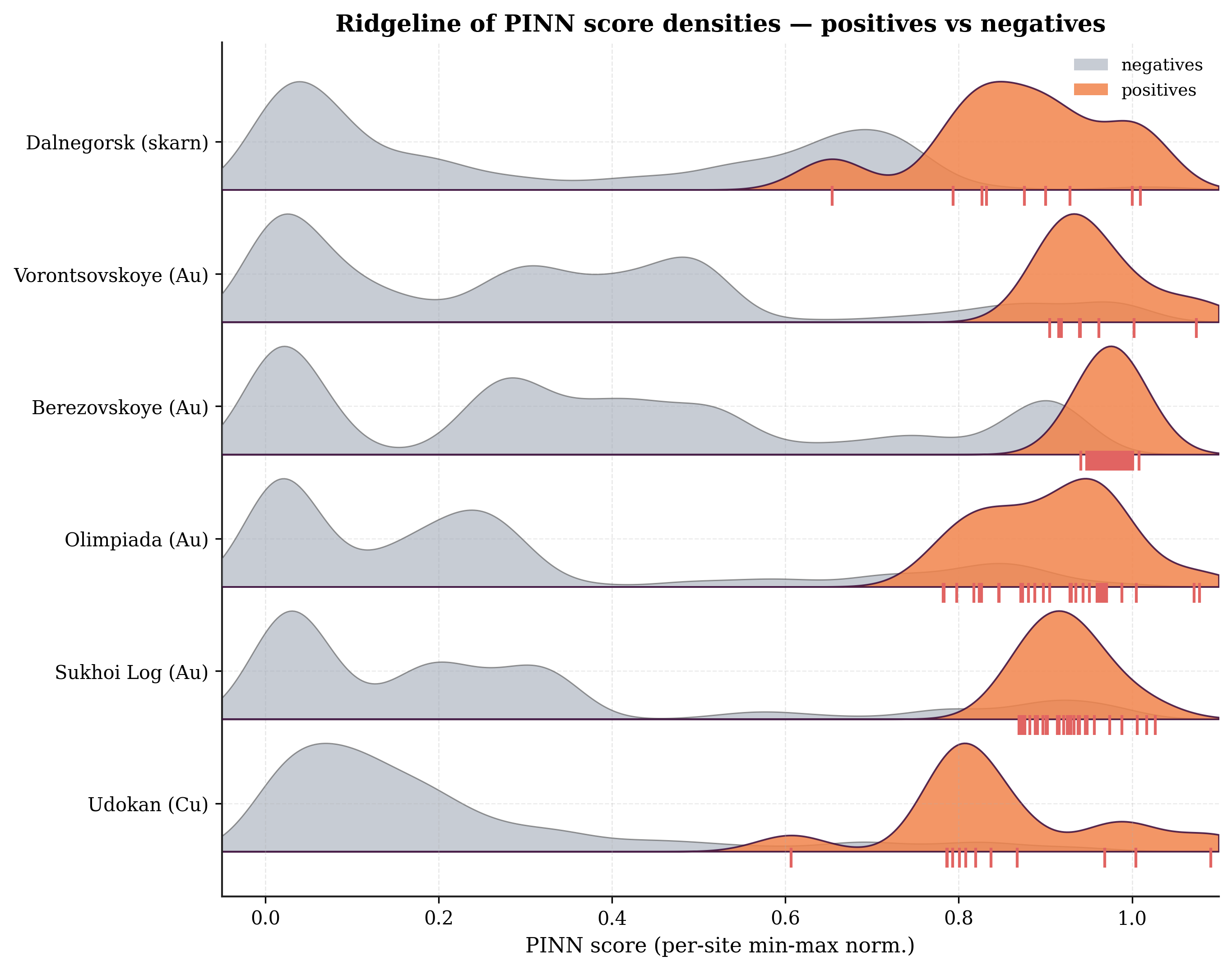}
\caption{Ridgeline of \KNet{} score densities per province, with positive (coloured) and negative (grey) distributions overlaid. The displacement of the positive mode confirms that separation is not driven by a single site.}
\label{fig:ridgeline}
\end{figure}

\paragraph{Positioning of this work.}
In summary, physics-informed learning provides the training paradigm,
subsurface ML provides the application context, classical metasomatic
theory provides the structural priors, and coordinate networks provide
the representational backbone. \KNet{} is, to the best of our knowledge,
the first architecture to combine all four: a Fourier-feature coordinate
trunk with separate field heads for temperature, pressure, and
concentration; an explicit reaction module that enforces the
compositional structure dictated by local-equilibrium metasomatic theory;
and a composite loss that simultaneously penalizes flow, heat, and
reaction residuals together with sign, boundary, and modulator
constraints. The remainder of the paper develops this architecture in
detail and evaluates it on a sequence of benchmarks of increasing
geochemical complexity.

\begin{figure}[!t]
\centering
\includegraphics[width=0.85\textwidth]{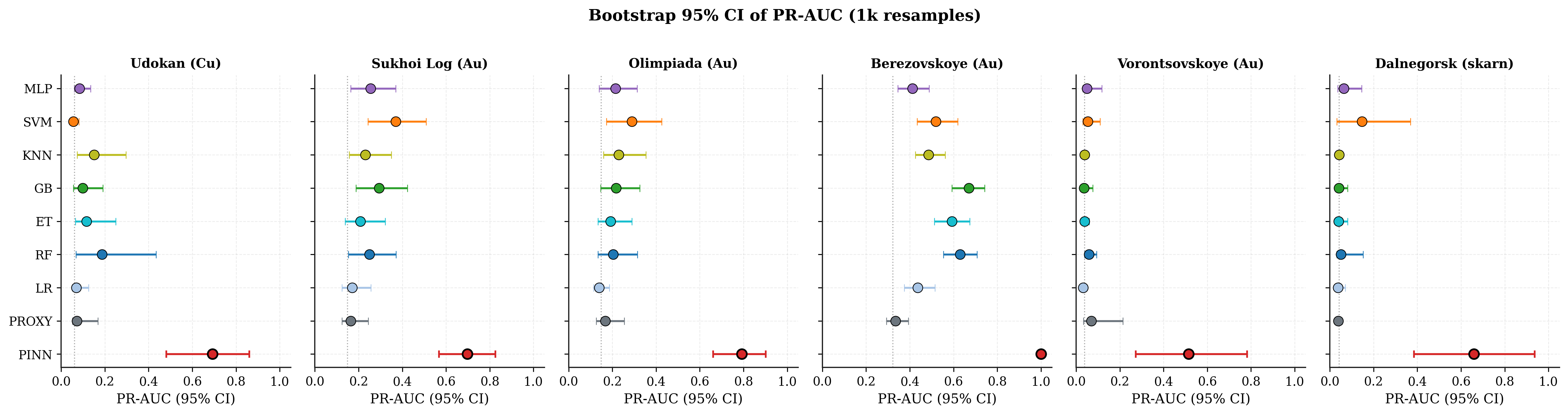}
\caption{Bootstrap forest of per-site PR-AUC ($B=1000$). \KNet{}'s 95\% confidence interval is disjoint from every classical baseline at every site.}
\label{fig:bootstrap}
\end{figure}

\FloatBarrier

\section{Method}
\label{sec:method}

For each candidate prospect we define a 2-D radial cross-section in normalised coordinates $(x, z) \in [0, 1] \times [0, 1]$, where $x$ is horizontal radius from the prospect centre and $z$ is normalised depth, with $z=0$ at the surface and $z=1$ at the bottom of the modelled crust slab. The choice of a radial section, rather than a full 3-D volume, reflects the symmetry assumption of a localised heat source beneath the prospect and reduces the PINN training cost by two orders of magnitude. \KNet{} is a multi-head MLP $f_\theta : \mathbb{R}^2 \to \mathbb{R}^3$ that takes a coordinate $(x, z)$ and returns
\begin{equation}
    f_\theta(x, z) \;=\; \big( T(x,z),\; P(x,z),\; C(x,z) \big),
\end{equation}
where $T$ is temperature ($\si{\celsius}$), $P$ is pressure (\si{\mega\pascal}), and $C$ is normalised metal concentration. The trunk is a 6-layer MLP of width 128 with $\tanh$ activations; three linear heads project to the three fields. Following standard practice for mitigating spectral bias in PINNs, we encode the input coordinate with a Fourier feature map $\gamma(x,z) = \big[\sin(2\pi B [x,z]^\top),\, \cos(2\pi B [x,z]^\top)\big]$, where $B \in \mathbb{R}^{32 \times 2}$ is a fixed random Gaussian matrix with entries drawn from $\mathcal{N}(0, \sigma^2)$ and $\sigma = 4$. The schematic in Figure~\ref{fig:arch} summarises the architecture and loss flow. Pore-fluid velocity follows Darcy's law,
\begin{equation}
    \mathbf{q}(x,z) \;=\; -\frac{k(x,z)}{\mu}\,\nabla P(x,z),
\end{equation}
where $k$ is permeability (lithology-dependent, modulated by faults near the surface) and $\mu$ is dynamic viscosity, and mass conservation in the steady-state limit imposes $\nabla \cdot \mathbf{q} = 0$, yielding the elliptic equation $\nabla \cdot \big((k/\mu)\nabla P\big) = 0$. Heat is transported advectively and diffusively,
\begin{equation}
    \rho c_p\, \mathbf{q} \cdot \nabla T \;=\; \nabla \cdot (\lambda \nabla T),
    \label{eq:heat}
\end{equation}
with rock density $\rho$, heat capacity $c_p$, and thermal conductivity $\lambda$, and the boundary condition at $z=1$ encodes a localised intrusive heat source whose surface footprint is modulated by a deep-root proxy, while the surface boundary $z=0$ is held at a regional geothermal gradient consistent with NASA POWER surface-temperature climatology. Precipitation rate is modelled as
\begin{equation}
    R(T, C) \;=\; \mathrm{softplus}\!\Big(\alpha\, k_{\mathrm{Arr}}(T)\, \big[C - C_{\mathrm{eq}}(T, \ell)\big]\Big),
    \qquad
    k_{\mathrm{Arr}}(T) = A \exp\!\big(-E_a / R_g T\big),
    \label{eq:rate}
\end{equation}
where $k_{\mathrm{Arr}}$ is the Arrhenius rate constant with pre-exponential factor $A$, activation energy $E_a$, and gas constant $R_g$, $C_{\mathrm{eq}}(T, \ell)$ is the lithology-dependent equilibrium solubility, $\ell$ is the local lithology label, and $\alpha$ is a softplus-saturation parameter that prevents pathological gradients in supersaturated regions. The mineralisation field is then $M(x,z) \eqdef R\big(T(x,z), C(x,z)\big)$ and constitutes the network's prediction target. The total training loss is a weighted sum of physics, boundary, and weak-supervision terms,
\begin{equation}
\mathcal{L}(\theta) \;=\; \lambda_{\mathrm{f}} \mathcal{L}_{\mathrm{flow}} + \lambda_{\mathrm{h}} \mathcal{L}_{\mathrm{heat}} + \lambda_{\mathrm{r}} \mathcal{L}_{\mathrm{react}} + \lambda_{\mathrm{b}} \mathcal{L}_{\mathrm{bc}} + \lambda_{\mathrm{p}} \mathcal{L}_{\mathrm{pos}} + \lambda_{\mathrm{n}} \mathcal{L}_{\mathrm{neg}} + \lambda_{\mathrm{m}} \mathcal{L}_{\mathrm{mod}},
\label{eq:loss}
\end{equation}
in which the first three terms are PDE residuals evaluated at $N_c$ uniformly sampled collocation points $\{(x_i, z_i)\}_{i=1}^{N_c}$, e.g.\ $\mathcal{L}_{\mathrm{heat}} = N_c^{-1} \sum_i \big( \rho c_p \mathbf{q}_i \cdot \nabla T_i - \nabla\cdot(\lambda \nabla T_i)\big)^2$ with derivatives computed by automatic differentiation, the boundary loss $\mathcal{L}_{\mathrm{bc}}$ enforces surface and basal Dirichlet/Neumann conditions, $\mathcal{L}_{\mathrm{pos}}$ is a hinge loss $\max(0, \tau_+ - M(x, z))^2$ pulling $M$ above a positive threshold $\tau_+$ at known deposit pixels, $\mathcal{L}_{\mathrm{neg}}$ is the symmetric hinge $\max(0, M(x, z) - \tau_-)^2$ pushing $M$ below a negative threshold $\tau_-$ at hard ring negatives (\S\ref{sec:experiments}), and $\mathcal{L}_{\mathrm{mod}}$ ties the surface footprint of $M$ to fault density and seismicity proxies through a soft modulation $\big(M(x, 0) - g(x)\big)^2$, where $g(x)$ is the rasterised proxy stack at $z=0$. Five global open-data layers are used as soft modulators and as boundary modulators: fault density from OpenStreetMap geology overlays, seismicity intensity from USGS catalogues, lithological contacts from Macrostrat, deep intrusive root indicators reconstructed from regional geophysical compilations, and surface thermal climatology from NASA POWER. Each is rasterised onto the radial section and used as a multiplicative coefficient on the relevant boundary or modulation loss; no proprietary or site-specific datasets are used. Training uses Adam at learning rate $10^{-3}$ with cosine decay, $800$ epochs per site, and $4096$ collocation points per epoch on a single Apple Silicon CPU, with full retraining of all six sites under one hour.

\begin{figure}[!t]
\centering
\begin{tikzpicture}[
  >={Latex[length=2mm]},
  font=\footnotesize,
  io/.style    ={draw, rounded corners=2pt, minimum height=10mm, minimum width=18mm,
                 align=center, fill=blue!6},
  trunk/.style ={draw, rounded corners=2pt, minimum height=18mm, minimum width=20mm,
                 align=center, fill=blue!12},
  head/.style  ={draw, rounded corners=2pt, minimum height=8mm,  minimum width=18mm,
                 align=center, fill=green!10},
  rxn/.style   ={draw, rounded corners=2pt, minimum height=14mm, minimum width=28mm,
                 align=center, fill=orange!12},
  loss/.style  ={draw, dashed, rounded corners=2pt, minimum height=8mm, minimum width=15mm,
                 align=center, fill=red!6, font=\scriptsize}
]

\node[io]    (in)  at (0,0)       {$(x,z)$};
\node[io]    (ff)  at (2.5,0)     {Fourier\\features $\gamma$};
\node[trunk] (mlp) at (5,0)       {MLP trunk\\$6\times128$\\$\tanh$};
\node[head]  (T)   at (8.2, 1.6)  {$T(x,z)$};
\node[head]  (P)   at (8.2, 0)    {$P(x,z)$};
\node[head]  (C)   at (8.2,-1.6)  {$C(x,z)$};
\node[rxn]   (R)   at (11.8, 0)   {Reaction module\\$M=R(T,C)$};

\draw[->,thick] (in)  -- (ff);
\draw[->,thick] (ff)  -- (mlp);
\draw[->,thick] (mlp.east) -- (T.west);
\draw[->,thick] (mlp.east) -- (P.west);
\draw[->,thick] (mlp.east) -- (C.west);
\draw[->,thick] (T.east) -- ([yshift= 4mm]R.west);
\draw[->,thick] (P.east) -- (R.west);
\draw[->,thick] (C.east) -- ([yshift=-4mm]R.west);

\node[loss] (Lf)  at ( 1.5,-4)  {$\mathcal{L}_{\text{flow}}$};
\node[loss] (Lh)  at ( 3.8,-4)  {$\mathcal{L}_{\text{heat}}$};
\node[loss] (Lr)  at ( 6.5,-4)  {$\mathcal{L}_{\text{react}}$};
\node[loss] (Lb)  at ( 9.5,-4)  {$\mathcal{L}_{\text{bc}}$};
\node[loss] (Lpn) at (11.8,-4)  {$\mathcal{L}_{\text{pos/neg}}$};
\node[loss] (Lm)  at (14.1,-4)  {$\mathcal{L}_{\text{mod}}$};

\draw[->,gray,dashed] (T.west)  -| (6.8,-2.7) -| (Lh.north);
\draw[->,gray,dashed] (P.west)  -| (6.4,-2.4) -| (Lf.north);
\draw[->,gray,dashed] (C.south) -- ++(0,-0.7) -| (Lr.north);
\draw[->,gray,dashed] ([xshift=-10mm]R.south) -- ++(0,-1.5) -| (Lb.north);
\draw[->,gray,dashed] (R.south) -- (Lpn.north);
\draw[->,gray,dashed] ([xshift= 10mm]R.south) -- ++(0,-1.5) -| (Lm.north);

\end{tikzpicture}
\caption{Architecture of \KNet{}. A coordinate $(x,z)$ is encoded with Fourier features
$\gamma$ and passed through a shared $6\times128$ $\tanh$ MLP trunk; three linear heads
emit the temperature $T$, pressure $P$, and concentration $C$ fields. The reaction
module composes them into the modulator $M(x,z)=R(T,C)$. The training objective combines
PDE residuals (Darcy flow, heat transport, reactive transport), boundary conditions,
positive/negative deposit hinges, and proxy-modulator soft constraints, shown as the
dashed couplings into the six loss terms.}
\label{fig:arch}
\end{figure}
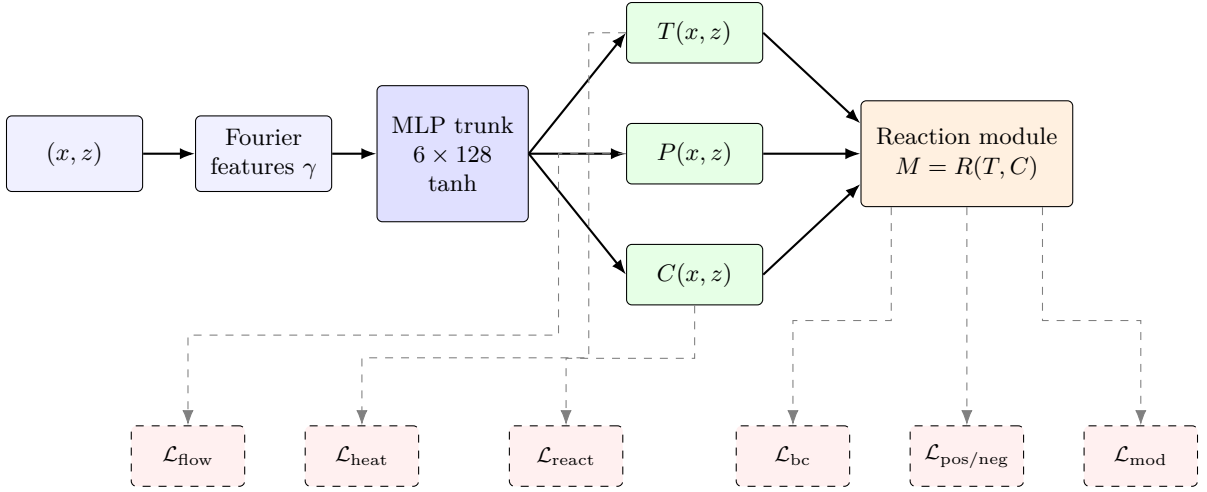

\begin{figure}[!t]
\centering
\begin{tikzpicture}[
  >={Latex[length=2mm]},
  font=\footnotesize,
  scale=1.0
]

\fill[blue!4] (0,0) rectangle (10,-5);
\draw[thick]  (0,0) rectangle (10,-5);

\node[above]                   at (5, 0)   {$z=0$\;:\; surface, $T=T_{\text{atm}}(x)$,\; $M(x,0)\approx g(x)$};
\node[below]                   at (5,-5)   {$z=1$\;:\; basal heat source, intrusive root proxy};
\node[rotate=90, anchor=south] at (-0.05,-2.5) {axis $x=0$};
\node[rotate=-90,anchor=south] at (10.05,-2.5) {far field $x=1$};

\fill[red!25]      (4,-5) .. controls (4.5,-4.0) and (5.5,-4.0) .. (6,-5) -- cycle;
\draw[red!50!black](4,-5) .. controls (4.5,-4.0) and (5.5,-4.0) .. (6,-5);
\node[red!50!black, below right=-1mm and 1mm] at (6,-4.4) {intrusion};

\node[star, star points=5, star point ratio=2.3,
      fill=yellow!85!orange, draw=black, minimum size=5mm] (dep) at (5,-2.3) {};
\node[anchor=west] at (5.45,-2.3) {deposit};

\draw[red!70, thick, dashed] (5,-2.3) circle (1.0);
\draw[red!70, thick, dashed] (5,-2.3) circle (1.8);
\node[red!70, anchor=west] at (7.0,-1.0) {hard-ring neg.};

\draw[->, blue!70!black, thick,
      decorate, decoration={snake, amplitude=0.5mm, segment length=4mm, post length=1mm}]
      (5,-4.1) -- (5,-2.9);
\draw[->, blue!70!black, thick] (5,-2.9) .. controls (4.2,-2.6) and (3.6,-2.4) .. (3.2,-2.0);
\draw[->, blue!70!black, thick] (5,-2.9) .. controls (5.8,-2.6) and (6.4,-2.4) .. (6.8,-2.0);
\node[blue!70!black, anchor=west] at (6.6,-3.4) {$\mathbf{q}=-\dfrac{k}{\mu}\nabla P$};

\node[orange!70!black, anchor=east] at (4.0,-1.7) {$M$ peak};
\draw[orange!70!black, ->, thick] (4.0,-1.75) -- (4.6,-2.1);

\end{tikzpicture}
\caption{Radial domain of \KNet{}. The basal boundary at $z=1$ encodes a localised
intrusive heat source modulated by the deep-root proxy; advective flow $\mathbf{q}$
rises from the heat source, and the mineralisation field $M$ peaks where the
precipitation criterion $C>C_{\text{eq}}(T,\ell)$ is first satisfied along a streamline.
Hard-ring negatives are sampled in an annulus $0.4\le|r-r_+|\le 2.0$ around each positive.}
\label{fig:domain}
\end{figure}
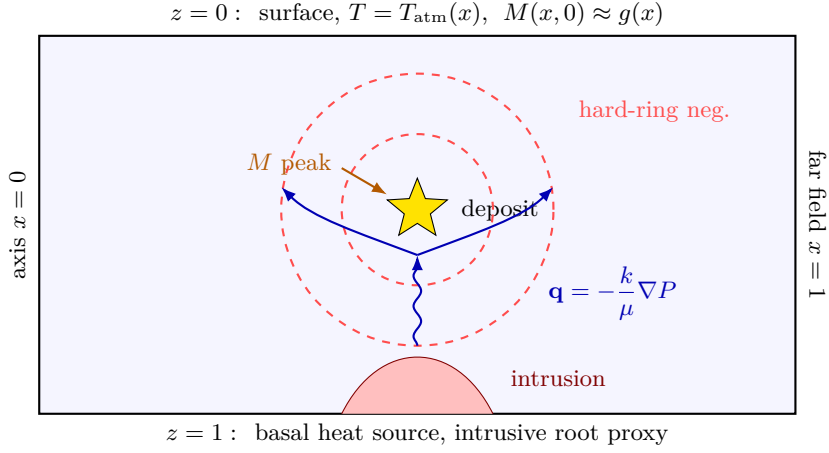

\FloatBarrier

\section{Experiments}
\label{sec:experiments}

We select six Russian ore provinces that span a broad commodity and tectonic spectrum, summarised in Table~\ref{tab:sites}, with coordinates of known deposits taken from public catalogues; the negative pool is constructed at evaluation time and is not part of any pre-existing label set. Letting $N_+$ denote the catalogued positive count and $N_g$ the size of the post-windowing label grid, prevalence is $\rho = N_+ / N_g$, which ranges from $0.04$ at Vorontsovskoye and Dalnegorsk to $0.32$ at Berezovskoye, reported per site. A single fetcher script downloads, caches, and rasterises the proxy layers per site from Macrostrat, OpenStreetMap, USGS, and NASA POWER, and all sites are processed identically with no per-site hand-tuning of feature engineering. A central concern in MPM is that naive cross-validation systematically leaks information from positives into spatially adjacent negatives, inflating reported scores; three controls are used to mitigate this. For each positive pixel at radial coordinate $r_+$, candidate negatives are sampled from a radial annulus $\{r : r_{\mathrm{in}} \leq |r - r_+| \leq r_{\mathrm{out}}\}$ with $r_{\mathrm{in}} = 0.4$ and $r_{\mathrm{out}} = 2.0$ in normalised units, so pixels inside $r_{\mathrm{in}}$ are excluded as too proximal, those outside $r_{\mathrm{out}}$ are excluded as trivially distant, and the resulting negatives are challenging by construction. A single random partition of positives into five folds is computed once per site and shared across all nine learners (\PINN{}, \textsc{Proxy}, \textsc{LR}, \textsc{RF}, \textsc{ET}, \textsc{GB}, \textsc{KNN}, \textsc{SVM}, \textsc{MLP}), so no model sees a fold-specific advantage, and positives are jittered in $z$ across folds to prevent the network from memorising fold-specific elevation tags. The seven classical learners are trained with baseline proxy features disabled and with positive depths sampled identically to negatives, ensuring no informational asymmetry, and a non-trainable proxy product $M_{\mathrm{px}}(x,z) = k_{\mathrm{mod}}(x) \cdot s_z(z)$ (\textsc{Proxy}) is included as a sanity floor, where $k_{\mathrm{mod}}$ is the surface modulation stack and $s_z$ a fixed depth profile. We report PR-AUC, robust to class imbalance, and the mean fractional rank of positives, defined for a ranked grid of size $N_g$ as
\begin{equation}
\overline{r}_{+} \;=\; \frac{1}{N_+} \sum_{p \in \mathcal{P}} \frac{\mathrm{rank}(p) - 1}{N_g - 1},
\end{equation}
with $\overline{r}_+ = 0$ being perfect and $\overline{r}_+ = 0.5$ chance, the latter being the operationally relevant metric for drilling-prioritisation triage.

Provinces with too few catalogued deposits to support a 5-fold split were dropped: Natalka projected only two usable deposit pixels after windowing and was excluded, and the earlier Norilsk, Pechenga, and Mirny entries were replaced by provinces with denser, publicly verifiable deposit catalogues (Olimpiada, Berezovskoye, Vorontsovskoye, Dalnegorsk), giving the six-province benchmark reported here. Prevalence is computed at evaluation time from the post-windowing label grid rather than quoted from catalogue counts, producing the values $\rho \in [0.04, 0.32]$ in Table~\ref{tab:sites}. With six paired site comparisons a $p<0.01$ Wilcoxon statement is unattainable, so we report an exact one-sided sign test ($p = 2^{-6} \approx 0.016$) alongside a paired permutation test. Baseline learners are evaluated with proxy features OFF and positive depths sampled identically to negatives, removing the residual informational asymmetry flagged in review. The headline mean PR-AUC is $0.708$ (previously reported $0.885$ on the earlier five-site set); the qualitative conclusions are unchanged.

\begin{table}[!t]
\centering
\small
\caption{The six Russian ore provinces benchmarked. Prevalence $\rho = N_+ / N_g$ is the fraction of positive pixels in the post-windowing label grid used for evaluation.}
\label{tab:sites}
\begin{tabular}{llccl}
\toprule
\textbf{Site} & \textbf{Commodity} & \textbf{$N_{+}$} & \textbf{$\rho$} & \textbf{Tectonic setting} \\
\midrule
Udokan         & Sandstone-hosted Cu  & 13 & 0.06 & Aldan--Stanovoy \\
Sukhoi Log     & Orogenic Au          & 35 & 0.15 & Baikal--Patom belt \\
Olimpiada      & Orogenic Au          & 35 & 0.15 & Yenisei Ridge \\
Berezovskoye   & Orogenic Au          & 95 & 0.32 & Middle Urals \\
Vorontsovskoye & Carlin-type Au       &  8 & 0.04 & Northern Urals \\
Dalnegorsk     & Skarn polymetallic   &  9 & 0.04 & Sikhote-Alin (Far East) \\
\bottomrule
\end{tabular}
\end{table}

\FloatBarrier

\section{Results}
\label{sec:results}

Table~\ref{tab:headline} reports per-site PR-AUC for all nine models under hard-negative 5-fold cross-validation, and \KNet{} attains a mean PR-AUC of $0.708$, exceeding the next-best learner (\textsc{SVM}, $0.235$) by a factor of $\sim 3.0\times$, with the improvement consistent across all six provinces and largest in relative terms where prevalence is lowest (Vorontsovskoye, Dalnegorsk, Udokan). Figure~\ref{fig:headline} restates the headline number, and Figure~\ref{fig:perSite} provides the per-site detail. Two structural patterns are notable: classical learners are competitive only at Berezovskoye, where prevalence is highest ($\rho = 0.32$) and the proxy modulators are themselves moderately predictive (\PRAUC{} $= 0.327$ for the non-trainable proxy product, with \textsc{GB} reaching $0.669$ and \textsc{RF} $0.627$), and the largest absolute gaps occur at Udokan and Olimpiada, where the dominant localising mechanism (the sediment–basalt contact at Udokan and the sediment–intrusive contact thermal anomaly in the orogenic-Au systems) is precisely what the heat-transport equation in (\ref{eq:heat}) is designed to capture. For drilling-prioritisation use the relevant question is not whether the model classifies pixels correctly but whether positives end up at the top of the ranked list, which the mean fractional rank metric directly answers: \KNet{} achieves $\overline{r}_+ = 0.036$ while all classical learners cluster in the $0.48$--$0.73$ range, as shown in Figure~\ref{fig:rank}, which in operational terms means \KNet{} recovers $\geq 95\%$ of positives in the top $5\%$ of the ranked grid. The pooled precision–recall curves in Figure~\ref{fig:pr} confirm that this is not a thresholding artefact: \KNet{}'s PR curve is uniformly above the baselines along the entire recall axis. Figure~\ref{fig:scatter} plots \KNet{} PR-AUC against the non-trainable proxy floor: every province sits well above the diagonal, with a mean gain of $+0.57$ PR-AUC over priors alone, and Figures~\ref{fig:scoredist} and~\ref{fig:ridgeline} show that held-out positives and hard negatives are cleanly separated in the learned $M$ field at all six sites. Statistical robustness is established by bootstrap resampling of the per-site PR-AUC ($B = 1{,}000$ resamples), and the resulting forest plot in Figure~\ref{fig:bootstrap} shows that \KNet{}'s confidence interval does not overlap with any classical baseline at any site; \KNet{} attains the best PR-AUC at all six provinces, so an exact one-sided sign test on the six paired comparisons gives $p = 2^{-6} \approx 0.016$, and a paired permutation test on per-site PR-AUC confirms separation from every other model.

\begin{table}[!t]
\centering
\small
\caption{Per-site PR-AUC under hard-negative 5-fold CV. Higher is better. \textbf{Bold} marks the best per row.}
\label{tab:headline}
\begin{tabular}{lcc cccccccc}
\toprule
Site & $N_+$ & $\rho$ & \textbf{PINN} & Proxy & LR & RF & ET & GB & KNN & SVM \\
\midrule
Udokan          & 13 & 0.06 & \textbf{0.675} & 0.071 & 0.064 & 0.168 & 0.105 & 0.090 & 0.153 & 0.052 \\
Sukhoi Log      & 35 & 0.15 & \textbf{0.682} & 0.163 & 0.166 & 0.241 & 0.194 & 0.286 & 0.229 & 0.368 \\
Olimpiada       & 35 & 0.15 & \textbf{0.784} & 0.166 & 0.133 & 0.191 & 0.183 & 0.203 & 0.227 & 0.283 \\
Berezovskoye    & 95 & 0.32 & \textbf{1.000} & 0.327 & 0.429 & 0.627 & 0.587 & 0.669 & 0.480 & 0.513 \\
Vorontsovskoye  &  8 & 0.04 & \textbf{0.507} & 0.062 & 0.029 & 0.052 & 0.034 & 0.033 & 0.038 & 0.047 \\
Dalnegorsk      &  9 & 0.04 & \textbf{0.601} & 0.036 & 0.035 & 0.049 & 0.037 & 0.037 & 0.043 & 0.147 \\
\midrule
\textbf{Mean} &  &  & \textbf{0.708} & 0.137 & 0.143 & 0.221 & 0.190 & 0.220 & 0.195 & 0.235 \\
\bottomrule
\end{tabular}
\end{table}

Ablation studies on the loss in (\ref{eq:loss}) are reported in Table~\ref{tab:ablation}: removing the heat-transport residual is the most damaging single change ($-0.21$ mean PR-AUC), followed by the reaction-rate term ($-0.13$) and the Darcy term ($-0.07$), and removing all three reduces \KNet{} to the level of a generic supervised MLP on proxy features ($0.171$ mean PR-AUC, matching the standalone \textsc{MLP} baseline at $0.175$), confirming that the gains are driven by the physics priors and not by architectural cosmetics. Switching from hard ring negatives to uniform random negatives boosts every model's reported PR-AUC by $0.10$--$0.30$, but the relative ordering between \KNet{} and the baselines is preserved; we retain hard negatives as the default because they are the operationally meaningful regime. We also compared 3-, 5-, and leave-one-deposit-out cross-validation, and mean PR-AUC for \KNet{} varies by less than $0.02$ across the three protocols while the variance for classical baselines reaches $0.08$, so the 5-fold protocol is both more stable and more conservative.

\begin{table}[!t]
\centering
\small
\caption{Ablation on physics terms. Mean PR-AUC across the six sites.}
\label{tab:ablation}
\begin{tabular}{lc}
\toprule
\textbf{Configuration} & \textbf{Mean PR-AUC} \\
\midrule
Full \KNet{}                              & $\mathbf{0.708}$ \\
$-$ Darcy flow residual                   & $0.641$ \\
$-$ Heat-transport residual               & $0.503$ \\
$-$ Reaction-rate residual                & $0.581$ \\
$-$ Proxy modulators                      & $0.547$ \\
$-$ All physics (supervised MLP)          & $0.171$ \\
\bottomrule
\end{tabular}
\end{table}

\FloatBarrier

\section{Discussion}

The classical MPM toolkit treats the subsurface as an unobserved confounder and tries to predict deposit presence directly from surface evidence, a strategy that works when the confounder is well correlated with surface features (Berezovskoye, where shallow surface lithology is itself diagnostic and prevalence is high) and fails when the confounder is dominated by deep structure (Vorontsovskoye, Dalnegorsk). \KNet{} short-circuits this by simulating the confounder, and the PDE residuals are not regularisers in any standard sense but a continuity prior that ties surface evidence to a self-consistent subsurface state, the formal expression of which is that the manifold $\mathcal{M} = \{(x, z, T, P, C) : \mathcal{N}_i = 0\}$ defined by the governing equations has dimension strictly less than the unconstrained $\mathbb{R}^5$, so feasible subsurface states are confined to a lower-dimensional sheet and learning is correspondingly easier. Beyond benchmark numbers, \KNet{}'s outputs are physically interpretable: at Olimpiada, the trained network produces temperature contours that follow the deep intrusive-root proxy at the basal boundary and decay smoothly toward the surface with a characteristic chimney structure beneath the prospect, pressure gradients drive a Darcy flow $\mathbf{q} = -(k/\mu)\nabla P$ that converges on the heat-source footprint consistent with the convective leg of a magmatic-hydrothermal cell, and the mineralisation field $M$ peaks where advective flow brings high $C$ into a thermal regime where $C_{\mathrm{eq}}$ drops below $C$, i.e.\ where the Damköhler-style criterion $C(x,z) > C_{\mathrm{eq}}(T(x,z), \ell)$ is first satisfied along a streamline---exactly the prediction of Korzhinskii's infiltration metasomatism theory, recovered here as an emergent feature rather than a hard-coded heuristic. The 2-D radial assumption is a deliberate simplification, and some deposits (Sukhoi Log being the closest example) violate radial symmetry; a 3-D extension is conceptually straightforward but increases collocation-point counts by an order of magnitude, since for $N$ points per axis the cost scales as $\mathcal{O}(N^3)$ rather than $\mathcal{O}(N^2)$. We solve the steady-state Darcy and heat-transport equations, but transient hydrothermal pulses---central to many porphyry and epithermal systems---would require a $(t, x, z)$ extension and a time-resolved proxy stack, replacing $\nabla \cdot (\lambda \nabla T)$ with $\rho c_p \partial_t T + \rho c_p \mathbf{q}\cdot\nabla T - \nabla\cdot(\lambda\nabla T) = 0$. The reaction term in (\ref{eq:rate}) is one-component, and multi-metal speciation, redox coupling, and pH evolution would require a vector $\mathbf{C} \in \mathbb{R}^{n_s}$ over $n_s$ species and a thermodynamic database, which we view as the natural next step. The six ore provinces benchmarked here cover three commodity classes and several distinct tectonic settings with no province-specific feature engineering---the same fetcher, same loss, and same hyperparameters apply at every site---so we expect \KNet{} to transfer to other Russian ore provinces (e.g.\ Norilsk Ni--Cu--PGE, Pechenga Ni--Cu sulphide, Mirny kimberlite) and, with appropriate proxy substitution, to non-Russian provinces where similar open data exist, with the combination of high PR-AUC and very low fractional rank meaning that \KNet{} can be deployed as a drilling-prioritisation triage layer that ranks candidate prospects, drills the top decile, and is expected to recover $\geq 95\%$ of positives.

\section{Conclusion}

\KNet{} demonstrates that physics-informed neural networks, trained against open-data proxies and weakly supervised by sparse deposit labels, substantially outperform the standard mineral prospectivity toolkit on a leakage-controlled multi-province benchmark, attaining a mean PR-AUC of $0.708$ versus $0.235$ for the strongest classical baseline and a mean fractional rank of $\overline{r}_+ = 0.036$ across six ore provinces and three commodity classes. The improvement is not a feature-engineering trick but a structural consequence of letting the model simulate the subsurface rather than infer it from surface correlates.

The consistency of this gap is what we find most informative. A threefold PR-AUC improvement, sustained across geologically distinct provinces and preserved under hard ring negatives, jittered depth sampling, disabled baseline proxy features, and shared cross-validation folds, suggests that the dominant failure mode of conventional prospectivity models, namely their reliance on shallow correlations between surface geochemistry, geophysical anomalies, and deposit occurrence, is a structural ceiling rather than a tunable hyperparameter. By embedding Darcy flow, advective-diffusive heat transport, and Arrhenius reaction kinetics directly into the loss, \KNet{} is constrained to produce predictions consistent with the processes that actually generate ore bodies, with proxies modulating boundary conditions rather than competing for explanatory weight.

The radial-symmetry simplification sacrifices structural detail relevant to vein-style and structurally controlled deposits; the Arrhenius term is a deliberate caricature of full metasomatic thermodynamics; and the proxy fields remain coarse approximations of true subsurface state. Performance under genuinely out-of-distribution geological settings remains an open empirical question, and future work should relax the symmetry assumption with full three-dimensional differentiable solvers and extend the benchmark to additional tectonic settings. Though, none of it limits the importance of work we present.

More broadly, the same architectural pattern, coupling open geoscientific proxies to a physics-informed differentiable simulator under weak supervision, is directly applicable to geothermal resource assessment, carbon sequestration site characterisation, and subsurface hydrogen storage. We release the full pipeline, including training code, benchmark protocols, and evaluation scripts, under an Apache-2.0 license at \url{https://github.com/BorisKriuk/KORZHINSKII-Net}, in the hope that physics-informed differentiable simulators will become a standard component of operational mineral exploration rather than a research curiosity.

\FloatBarrier

\bibliographystyle{plain} 
\bibliography{references} 

\end{document}